\documentstyle[preprint,epsfig,aps]{revtex}
\draft
\tightenlines
\begin{document}
\title{Nematic - Isotropic Transition in Porous Media - a Monte Carlo Study}
\author{K. Venu\cite{venu}, 
        V. S. S. Sastry\cite{sastri} 
        and K. P. N. Murthy\cite{kpn}}
\address{\cite{venu}${}^{,}$\cite{sastri} School of Physics, University of Hyderabad, 
                 Hyderabad 500046 Andhra Pradesh, India\\
        \cite{kpn} Materials Science Division, 
           Indira Gandhi Centre for Atomic Research,\\
         Kalpakkam 603 102, Tamilnadu, India}
\date{\today} 
\maketitle
\begin{abstract}
We propose a lattice model to simulate  the influence 
of the porous medium on the Nematic-Isotropic phase 
transition of liquid crystals confined to its pores.
The effects of the pore surface and  the pore interconnectivity 
are modelled through a disorder parameter.
Monte Carlo 
calculations based on  the  model 
leads to results that compare  well 
with experiments on  the suppression of the 
transition temperature with decreasing pore size, the 
softening of the transition from the first order to continuous 
when the 
pore size becomes small and  the enthalpy of transition.
\end{abstract}
Statistical mechanics  of liquid crystals 
in confined geometry is  a subject of  interest
both from technological and basic science points of view 
 \cite{tandf1,pcz}.   
Many electro-optical devices operate  based on the 
properties of liquid crystals  confined in suitable geometries. 
For example, incorporating liquid crystal in gel network has led to 
a new type of nematic  display device with a significantly 
improved switching property \cite{hz}. The 
academic interest in this subject 
stems  from the fact that 
confinement leads to several complex, unexpected and 
interesting phenomena that are not very well understood. 
The basic issue concerns the phase behaviour of the 
confined liquid crystals which 
is determined  by the  delicate balance established 
amongst the competing effects of the disorder induced by the 
confining surface, the finite size,  the orientational 
order and  the temperature.  
A porous medium with a random 
network of interconnected pores of random sizes and shapes  
provides a simple confining medium 
\cite{tandf1,pcz}.
The sizes of the  pores in the 
medium,  pore surface roughness,  
 pore connectivity {\it etc.,} influence 
the nature and content of the  phase  and its transition. 

In this letter we shall  focus attention on the 
Nematic-Isotropic (N-I) phase transition. Experiments tell us that  
the transition  temperature, denoted by $T_{{\rm NI}}$, is lower in
the porous  medium as compared to the bulk sample; also $T_{{\rm NI}}$ 
decreases with decreasing pore size \cite {wzgbs}.  
The N-I transition is weakly first order in the bulk. In the porous 
medium, however, the transition softens as the pore size
decreases and becomes continuous for small pores \cite{gzad}.  

Several models have been proposed and investigated to understand the 
above  experimental findings. Our interest here is to study 
the orientational ordering of the liquid 
crystals; hence lattice models  with 
nearest neighbour interaction would suffice. 
There are numerous approaches that have been tried out to model
the influence of the porous matrix. 
These include explicit 
construction of an independent pore with different
 rigid boundary conditions and 
director configurations
\cite{bp2,bp1},  Potts spin models with the porous medium approximated
by the diffusion-limited cluster-cluster aggregate \cite{uhj},
random field Ising models \cite{potts}, 
random anisotropy nematic
models \cite{cksa} and  
models based on dilution \cite{dilution1,dilution2}.
An important objective of all these models has been 
to make a quantitative prediction of the 
depression of $T_{{\rm NI}}$ with decreasing 
pore size. The  models 
proposed so far have been 
successful only qualitatively:   
the predicted decrease  of $T_{{\rm NI}}$ with decreasing pore size 
is atleast 
an order of magnitude higher, see {\it e.g.} \cite{uhj},  than 
what has been observed experimentally.  
Such a  large discrepancy  
has been  attributed to several reasons. For example, 
in the independent pore models, the influence 
of the interconnectivity of the pores is not accounted for \cite{pcz}. 
In the dilution models, only the excluded volume effect 
of the pore surface is modelled. 
Despite this,  the  inter-particle interaction seems to be 
excessively perturbed in these  models \cite{dilution2}.
 
In this letter, we propose a model that takes into account the 
effect of the disorder, induced by the 
pore surface as well as pore connectivity.
Consider an $L\times L\times L$  
cubic lattice with a total of $L^3$ lattice sites.   
Each lattice site holds a liquid crystal molecule interacting only 
with its six nearest neighbours as per 
the angular part of the Berne - Pechuka (BP) 
potential \cite{dilution2}. Periodic boundary conditions are imposed. 
For a given value of $L$ we 
select  a certain number,  $N_d $, 
of lattice sites randomly and declare the liquid crystal molecules 
on these sites to have quenched random orientations. 
This disorder parameter $N_d$ is key to our model. It is  
defined in such a way that it serves as a 
good measure of the confining effects of  a single pore  as well
as the  connectivity of the random network of pores, {\it wide infra}.

Consider a pore that contains $N$ liquid crystal 
molecules. Of these, on the average, 
$N^{2/3}$ molecules will be on the surface and 
these are the ones that are directly 
influenced by the pore surface.
In a lattice model,  
the substrate site perturbs 
one out of the six nearest 
neighbour interactions ({\it bonds}).
However, in our model, we propose to  take care of 
the surface effect by 
placing molecules with 
quenched random orientations inside the 
lattice; each such {\it quenched} molecule 
has  six bonds.
Thus it is reasonable
to say that  the relevant number 
of randomly quenched molecules required to represent
the surface effects of a 
single pore should be  
$6^{-1}N^{2/3}$.  
Next,  consider 
the  cubic lattice 
whose  volume is $L^3 v$, where $v$ is the volume of a 
single
liquid  
crystal molecule.
If  the typical volume of a pore is $V$, then 
the number of pores in the lattice 
cube  is $L^3 v /V$. It immediately follows, $N_d = L^3 v V^{-1} 
N^{2/3} 6^{-1}$.  
This can be expressed conveniently in terms of  the 
typical size (diameter)
of a pore ($L_p$), the linear size of the lattice system ($L$)  and  the 
volume of a single liquid crystal molecule  $(v)$ as,  
\begin{equation}\label{dl}
N_d (L, L_p , v) = {{L^3}\over{L_p}} 
\left( {{v}\over{36\pi }}\right)^{1/3} \ .
\end{equation}
It is clear that a set of  quenched sites represents
the effect of the wall of a single pore. Several such sets of
 quenched sites are
distributed independently, randomly and uniformly in the lattice cube and 
they represent the effect of the walls of all the pores. 
Each active lattice site 
is connected to every other ones through a path consisting
of  nearest neighbour connecting segments. In this sense our model
represents, approximately, the interconnected network of pores.
Also, we can conveniently employ periodic
boundary conditions which is known to have least finite size effects.
In addition, wherever possible, we can employ finite size scaling  to
calculate the macroscopic properties in the thermodynamic limit.
For a typical liquid crystal molecule of 8CB,  
studied extensively in different porous media,
$v=480$ \AA $^3$. The porous media studied experimentally 
have typical sizes ranging from $80$\AA\ to $400$\AA\ . This situation
can be simulated by making proper choices of $L_p$ and $v$. 

Monte Carlo simulations with standard 
Metropolis algorithm   
were carried out for   
five pore sizes (approximately corresponding to experimental situations
involving 8CB molecules) 
and for each pore, for five values 
of $L$. The values of pore sizes considered 
in our study are $L_p =400,\ 180,\ 150,\ 110,\ 
$ and $80$ \AA\ , besides the bulk liquid crystal, which corresponds 
to $N_d = 0$ . For each case, 
we considered system sizes of $L=10,\ 
15,\ 17,\ 20,$ and $22$,  
and simulations were carried out 
at forty values of $T$, bracketing the 
transition temperature. Also, the simulations  
were repeated for three random disorder configurations
and the results were averaged over these runs.  
For each case, the Monte Carlo simulations 
were carried out in three stages.
In the first stage,  short runs of  
$20,000$ to $50,000$ Monte Carlo (MC) cycles were carried out.  
From the data generated,  we calculated the values 
of  the specific heat $C_V$ at
the chosen temperatures. 
The transition temperature was located approximately 
as the one at which $C_V$ is maximum.  
In the second stage, extensive simulations 
consisting of $500$ thousand  MC  cycles
with initial hundred thousand cycles 
ignored for equilibration purpose,  were 
carried out at the  
approximate transition temperature estimated  
in the first stage. From these results, 
we calculated the distribution of energy $E$. 
Then, employing 
Ferrenberg-Swendson  reweighting technique
 \cite{fs}, the energy distribution was  
analyzed to yield fresh specific 
heat profiles as a function of $T$. 
The location of the specific heat peak at this 
stage yields the next level of approximation 
to the transition temperature. 
In the third stage,  large Monte Carlo 
runs of five hundred thousand MC cycles 
were carried out  at the 
transition temperature estimated in the second stage
  and this process 
was repeated for three disorder configurations. 
The distribution of energy 
thus obtained 
was analyzed for each case, employing again  
Ferrenberg-Swendson reweighting 
technique to obtain profiles of specific heat 
$C_V$, ordering susceptibility $\chi$, and 
the fourth cumulant (also known as Binder 
cumulant)  \cite{binder}, $V_4$ of energy  fluctuations.

$T_{{\rm NI}}$ deduced from 
the data on $C_V$ as well as from that 
on $V_4$  are plotted 
as function of $L^{-3}$ in Fig. 1(a), 
for the bulk sample.  
The two sets 
of data fall on straight lines establishing  
finite size scaling. 
Extrapolating the system size to infinity, 
we get the transition
temperature in the thermodynamic limit.
Similar scaling behaviour was found for the 
simulations carried out for a pore size of
$400$ \AA\  , and is shown in Fig. 1(b). 
For smaller 
pores 
however, we could not get 
a neat finite size 
scaling behaviour;
hence, for these cases,  we have calculated  the
transition temperature by taking the average 
of $T_{{\rm NI}}$
obtained from $C_V$ and $V_4$ for 
the largest lattice size 
considered ($L=22$). 
The data on $T_{{\rm NI}}$ 
thus obtained and normalized to the 
bulk transition temperature, is plotted as a 
function of the inverse of  the pore size ($L_p$),    
in  Fig. 2, alongwith 
the experimentally measured 
data \cite{bp2}.
We find that our 
results match remarkably well with the measured 
values. For comparison, we have shown the 
coresponding predictions of the independent 
pore model, in the same figure.  

A set of macroscopic quantities that would reveal 
the nature of the phase transition through their 
finite size scaling patterns, includes the peak 
amplitude and Full Width at Half Maximum (FWHM) of the 
temperature profiles of $C_V$, $\chi$ and $V_4$. 
We have shown in Fig. 3 (Left)  the 
peak amplititude of the 
specific heat $C_V$ versus $L^3$ for $L_p =0$ (bulk sample) and   
$400$\AA\ .  
In the case of bulk and pore size 
$L_p = 400$ \AA\  , the peak amplitude of 
$C_V$
scales as $L^3$, as expected for first order transition.
For $L_p = 180$\AA\    
however we find 
$L^3$-scaling does not hold good \cite{vsm}.  
This failure of $L^3$- scaling 
is an indication of the softening of the phase transition   
as the pore size decreases. Similar results on the 
orientational susceptibity are shown in Fig. 3 (Right). We find that 
for the bulk and pore size $400$\AA\ , the susceptibility scales 
as $L^3$; for pore size $180$\AA\ the scaling breaks down \cite{vsm}.  

We  calculated the  minimum of $V_4$ as a function of $L$ for 
the bulk as well as for pore size $L_p = 400$ \AA\  . 
In the asymptotic ($L\to\infty$)  limit,  
this value reaches $0.64$. However for the case
with  
$180$\AA\ 
we find that asymptotically the minimum of  
$V_4$ reaches a value of  $0.66$,  expected in a 
continuous  transition. 
This gives a further indication  of the cross over 
of the transition from the  first to the continuous one  with decreasing 
pore size, for details see \cite{vsm}.

We  have analyzed  the data on the FWHM 
of the temperature profiles of $C_V$, $\chi$ and $V_4$.
For  first order phase transition, these 
quantities are expected to scale as $L^{-3}$. Indeed we find 
such a scaling behaviour for the bulk and 
for pore size of $L_p =400$ \AA\  . 
The results for 
the pore size of $L_p=180$ \AA\ ,  
do not exhibit such a scaling  
giving further evidence for the cross over phenomenon. 
The details of these results are given in \cite{vsm}.
 
The energy distribution near a first order 
transition should show a double peak structure
corresponding to a double well potential. 
This corresponds to the coexistance of the two phases
separated by a potential barrier, representing 
surface energy terms arising from domain formation.
 Employing reweighting 
techniques, the energy distribution obtained as described 
earlier, was extrapolated 
to yield distributions in the neighbourhood 
of the transition temperature.
We found that in the case of bulk as well as for 
pore of size $400$  \AA\  , a double peak structure 
in the distribution of energy  yielding  
a profile with equal peak heights.
The energy 
distribution thus obtained, was transformed to a quantity
which differs from free energy only by an 
additive constant. The 'free energy' thus obtained had a 
double well structure for the bulk sample. Also the barrier 
height was found to increase with increase of $L$.  
A similar result 
was  obtained for the 
pore size $400$\AA\  . We then plotted the barrier height as a function
of $L^2$ for the bulk sample as well as for $L_P=400$\AA\ . The curve 
was found to be linear establishing  $L^2$-scaling. This confirms that the
transition is first order. 
We 
could not extract such  structures  (typical for  
first order transition)  for the cases with 
pore sizes $180$\AA\   and less. 
This provides a definitive evidence that 
the nature of the transition indeed  
crosses over from the first order to the continuous one,  when the 
pore size  decreases.  
The details 
are given  
in \cite{vsm}. 
  
The enthalpy exchanged in the transition can be obtained 
as integrated area of the specific heat profile and 
is depicted 
in Fig. 4 as a function of pore size; also shown 
are 
the experimental
results. The agreeement is found to be  very  good. 

In conclusion, we have proposed a  model to study the phase behaviour 
of liquid crystals in a porous medium. 
We have given a method
for calculating the disorder parameter $N_d$ in terms of the 
typical pore size, the 
system size, and the volume of the liquid 
crystal molecule. 
The predictions of our model 
on the suppression of the 
transition temperature with decreasing 
pore size, on the softening of the  transition  as
the pore size decreases  and on the enthalpy of transition 
compare well with experimental findings. 

We thank G. Bharadwajkumar and 
V. Satheesh for help  during  the early stage of the  calculations, 
B. V. R. Tata and M. C. Valsakumar
for discussions and     
the  Board of Research in Nuclear Sciences, India
(Project No. 99/37/09/BRNS)  
and  the office of the Naval Research, USA  
(Grant No. N00014-97-1-0994) for support.

\newpage
{\centerline{\bf \large FIGURE CAPTIONS}}

Figure 1    Transition temperature deduced from Specific heat ($\times$)  and the Binder
            Cumulant ( o ), plotted against $L^{-3}$. (a) for the bulk (b) for $L_p = 400$ \AA\ .
            The error bars are also  marked.

Figure 2    $T_{{\rm NI}}(L_p)/T_{{\rm NI}}$(bulk) versus $1/L_p$.
            o - Experiment; + - present simulations;
            $\times$  - independent pore model.

Figure 3    Maximum of   specific heat (Left) and   of  the orientational
             susceptibility (Right) plotted against $L^{3}$.  $\times$ - bulk; o - pore size 400\AA\ .
    
Figure 4   Variation of relative enthalpy with pore size;
           o - experimental values; $\times$  - our results .  
\newpage
\begin{figure}[ht]
\centerline{\psfig{figure=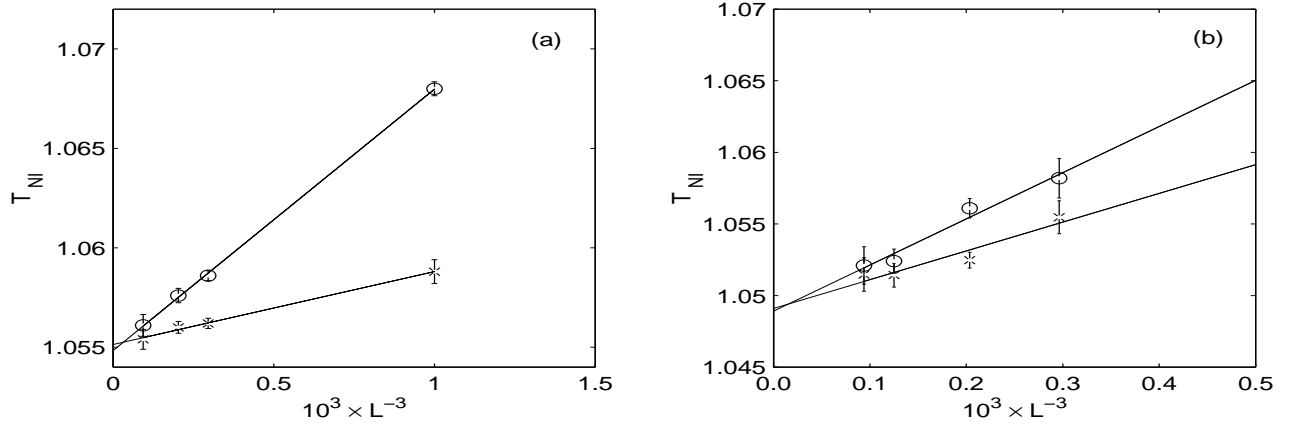,height=7.0cm,width=17.0cm}}
\caption{\protect\small 
  Transition temperature deduced from Specific heat ($\times$)  and the Binder
      Cumulant ( o ), plotted against $L^{-3}$. (a) for the bulk (b) for $L_p = 400$ \AA\ .
      The error bars are also  marked.
}
\label{fig1}
\end{figure}
\newpage
\begin{figure}[ht]
\centerline{\psfig{figure=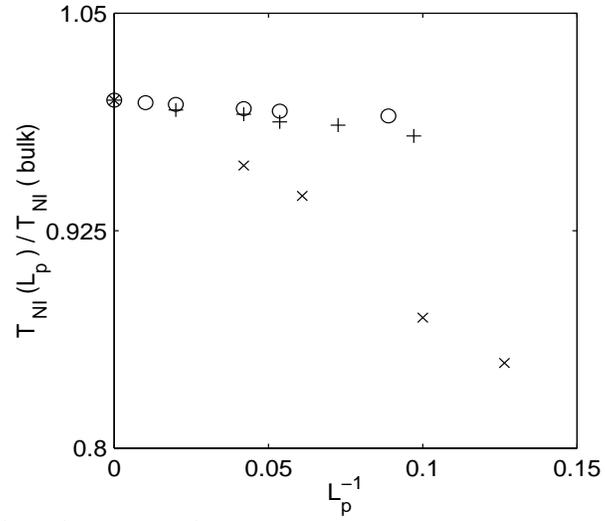,height=7cm,width=8.0cm}}
\caption{\protect\small  
$T_{{\rm NI}}(L_p)/T_{{\rm NI}}$(bulk) versus $1/L_p$. 
o - Experiment; + - present simulations;
$\times$  - independent pore model
}
\label{fig2}
\end{figure}
\newpage
\begin{figure}[ht]
\centerline{\psfig{figure=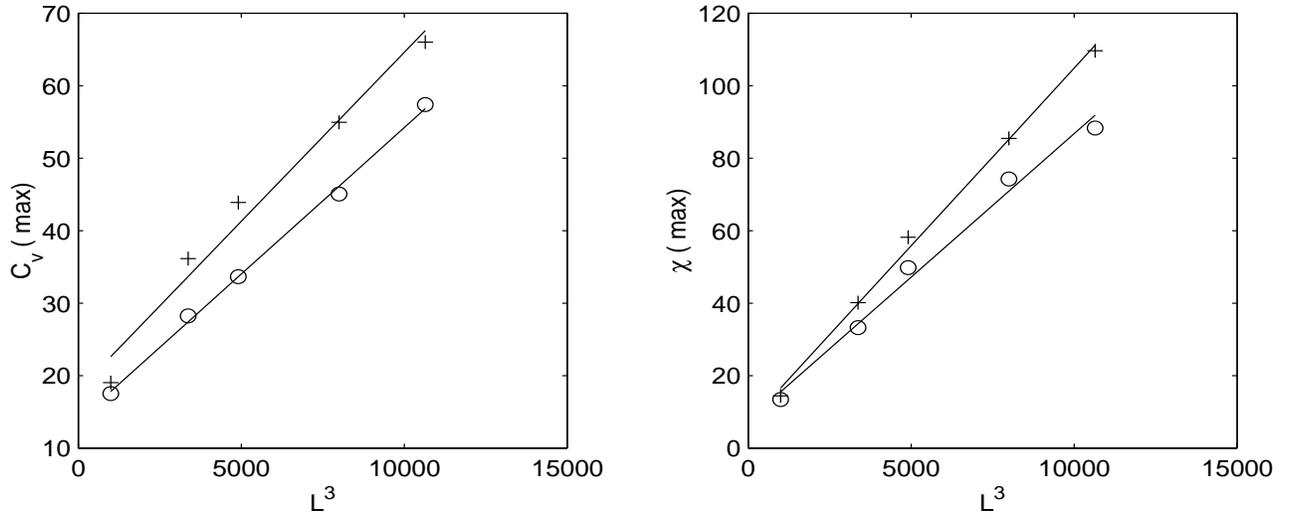,height=7.0cm,width=17.0cm}}
\caption{\protect\small 
 Maximum of   specific heat (Left) and   of  the orientational
susceptibility (Right) plotted against $L^{3}$.  $\times$ - bulk; o - pore size 400\AA\ .
 }
\label{fig3}
\end{figure}
\newpage
\begin{figure}[ht]
\centerline{\psfig{figure=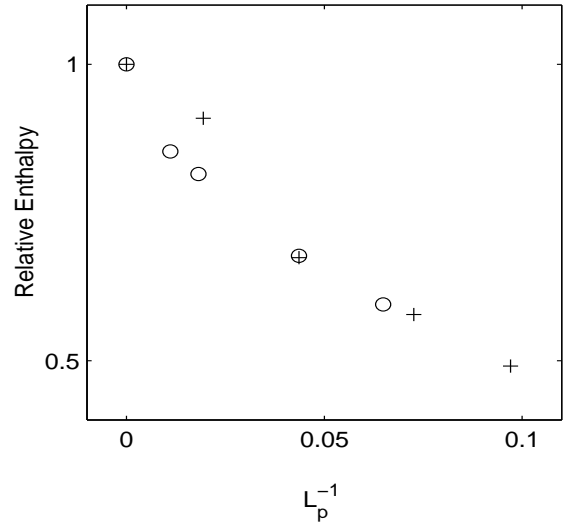,height=7cm,width=7.5cm}}
\caption{\protect\small  
Variation of relative enthalpy with pore size;
o - experimental values; $\times$  - our results.}
\label{fig4}
\end{figure}
\end{document}